# Mediatised Participation: Citizen Journalism and the Decline in User-Generated Content in Online News Media

**Simón Peña-Fernández[1]*, Ainara Larrondo-Ureta[2] and Irati Agirreazkuenaga**

1. University of the Basque Country (UPV/EHU) (ROR: 000xsnr85)
   simon.pena@ehu.eus / https://orcid.org/0000-0003-2080-3241
2. University of the Basque Country (UPV/EHU) (ROR: 000xsnr85)
   ainara.larrondo@ehu.eus / https://orcid.org/0000-0003-3303-4330
3. University of the Basque Country (UPV/EHU) (ROR: 000xsnr85)
   Irati.agirreazkuenaga@ehu.eus / https://orcid.org/0000-0002-4258-7714

* Corresponding author

**Abstract**: The second generation of web tools shook the journalist profession approximately two decades ago with the proactive incorporation of audiences into the media. Citizen journalism and user-generated content arose as an object of interest due to the democratising value of participation attributed to them, with empowered citizens who could emulate the professional and institutional practises of journalists. However, difficulties soon came to the surface, and audience participation in news media began to be limited. Within this context, this article conducts a critical review of studies on audience participation in news media based on a systematic literature review. The results indicate that, in general, audiences showed low interest in the creation of informative content and that their participation has grown increasingly problematic. In addition, journalists are reticent as they defend their professional role above all else, while company strategies have prioritised making participation profitable. For this reason, the idea of citizen journalism that offers user-created content through the media appears to be a thing of the past, with many characteristics that could define it as a failed innovation. Therefore, the text concludes that audience participation in the media could be defined as mediatised participation.

## 1. Introduction

In the changing professional landscape of journalism, new social actors coexist with established, organisational and institutional actors that produce, share, and distribute news in an increasingly complex hybrid media system (Chadwick 2013; Belair-Gagnon et al. 2019).

Thus, the second generation of web tools (commonly known as web 2.0) shook the journalist profession approximately two decades ago with the proactive incorporation of audiences into journalism. Initial approaches (Bowman and Willis 2003; Gilmor 2004) were frequently formulated as veritable news utopias (Mosco 2004) that defended the democratising value of participation, with empowered citizens who could emulate the institutional and professional practises of journalists (Boccia-Artieri 2012) and who would create messages capable of competing with those created by the news media in their day-to-day lives (Rodriguez 2001; Atton 2001). Since then, citizen journalism and audience participation in the media has been a recurring subject of study (Domingo et al. 2008; Paulussen and Ugille 2008; Allan and Thorsen 2009; Borger et al. 2013; Peña-Fernández et al. 2019).

During these initial turn-of-the-century years, fascination with participatory journalism and the reversal of roles in the news model was partially due to economic and credibility issues affecting the media (Deuze et al. 2007). This was not merely a technological shift; rather, it was a new cultural phenomenon that redefined traditional news spaces in an area where the limits became more and more blurred (Lewis 2012). In a digital economy in which the distribution and consumption of goods and services are increasingly based on cooperative relations (Ramella and Manzo 2020), the future is opening towards interactive and connective news production, where users and media would coexist, collaborate, and also compete, in the shared task of building reality (Deuze 2009).

The initial enthusiastic approaches, which argued that these innovations would inevitably and automatically cause social change, were largely driven by idealism and technological determinism (Larrondo-Ureta et al. 2023). However, it must be remembered that all innovations, in order to be considered successful, require not only achieving a technical advance but also social appropriation of them, that is, that citizens use them naturally in their lives (Peña-Fernández et al. 2019).

Indeed, it was this social appropriation where audience participation and citizen journalism ran aground. While the idea of participation as a driving force of democracy understood that institutions were adaptable and that audiences could mobilise if given the opportunity (Deuze et al. 2007), almost from the very beginning, we could verify that the news media was ill-predisposed to cede control of the editorial process to audiences and that these audiences also displayed a limited desire to participate in the news-creation process (Hermida and Thurman 2008; Singer et al. 2011; Hermans et al. 2014).

In recent years, this initial optimism has been cut short. The most usual trend in studies on participation leans towards problematisation (Quandt 2018), the distrust it creates amongst media professionals, and the sales focuses of news companies.

Within this context, this article conducts a critical review of studies on audience participation in news media based on a systematic literature review. The objective is to define this audience incorporation and the limits found based on the analysis of the three main actors in the process: the audiences themselves; the professionals; and the news companies.

## 2. Materials and Methods

To draw up this article, we conducted a systematic literature review (Grant and Booth 2009; Codina 2020a) based on the concepts of “citizen journalism” and “participatory journalism”, as well as other similar definitions (journalism 2.0, open-source journalism, etc.). As a time range, the period spanning from 2000 until 2022 was chosen. The search encompassed main academic databases (Web of Science, Scopus) and was completed with consultations in the Google Scholar, EBSCO, and ProQuest search engines.

After the systematic search, the results obtained both from quantitative data (number of references obtained) and from qualitative data (topics addressed in abstracts and keywords) were categorised, following the indications of the panoramic review technique or scoping review with PRISMA (Codina 2020b).

As Tricco et al. (2018) explain, scoping reviews are a type of knowledge synthesis that follows a systematic approach to map evidence on a topic and identify main concepts, theories, sources, and knowledge gaps. Likewise, these types of reviews are useful to evaluate the characteristics, variety, and volume of research on a topic and allow us to summarise findings from a diverse range of studies that may vary in methods or disciplines (Tricco et al. 2018).

Following the mentioned search criteria, the first sample was composed of 972 documents. After applying the criteria of adequacy to the topic, time range, type of text (article or book chapter) and language, the reviewed selection consisted of 437 documents. Finally, for the selection of the final sample, empirical studies and research with the greatest academic impact were prioritised. This text synthesises the contributions of 125 of these texts.

## 3. Results

### *3.1. Participation and Interaction*

In the beginning, the attention drawn by participation in the media was closely linked to its conception as an intrinsically democratic tool. This concept was used in a highly diverse fashion, although Carpentier (2016); Carpentier et al. (2019) identifies two main trends into which these definitions can be grouped: one with a more inclusive nature and more sociological profile, where participation is understood as any of the methods of social interaction; and another that is more selective and nearer political studies, where participation is understood as a way of sharing power.

If we extend these two meanings, we understand that, in the broadest definition, participation is the equivalent of interaction, meaning all the ways in which the audience can interact with the media and its content, from content selection to participation in a survey or sharing information. The broadest definitions of citizen journalism (Goode 2009; García-De-Torres 2010) lean towards this definition. However, if we view participation as a balance in the shared exercise of power, it would be limited to productive interactivity (Rost 2006), meaning the audiences' capacity to create content and actively participate in the news-creation process in the media (Deuze 2006; Robinson and Wang 2018).

Unlike other forms of civic participation, the creation of this type of content would be understood as participation in the media and not participation through the media (Carpentier 2011; Ahva 2017), which would imply a purpose other than news creation. There would be a differentiation between the technology that allows users to control and personalise content and the platforms that allow citizens to create and distribute content in the public discourse or integrated structural participation (Peters and Witschge 2014).

In other words, participation could be understood as the most advanced way to interact with the media, where audiences take on roles usually reserved for journalists, while not all forms of interaction could be considered participation.

*3.2. Audiences' Lack of Interest and Problematic Participation*

Although more limited than today, studies on participation in the media have a long history (Mendiguren-Galdospin and Canga-Larequi 2017; Marzal-Felici et al. 2021). When audiences were empowered, this was the beginning of a new stage, when they began offering users channels by means of which to make their voices heard.

However, it soon became clear that one of the main obstacles to the development of citizen journalism was that, in general, audiences displayed very little interest in the creation of news content (Lowrey and Anderson 2005; Chung 2008; Karlsson et al. 2015). Those who showed interest were driven by the individual defence of certain ideas so that they would reach the public sphere or by dissatisfaction with traditional news media and the ideas that they defended (Larsson 2014).

Notwithstanding, in global terms, the silence of audiences in creating news content has been deafening (Milioni et al. 2012; Masip 2015). Even though the media generally offered little space to create news content (Meso-Ayerdi et al. 2014; Pantic 2018), even in spaces where they were given a place, very few of these contents ended up prospering or enjoying noteworthy coverage (Scott et al. 2015). Citizens were more interested in creating content related to popular culture or their daily lives than journalistic content of general interest (Jönsson and Örnebring 2011; Xiang 2019).

Along with a lack of motivation, there was also a lack of professional resources held by those who created this content. Concern over the weak content created by the users (Meso-Ayerdi 2013) includes their limited access to information sources (Reich 2008) and the absence of professional training (Kus et al. 2017). These issues can lead to them incorrectly reproducing professional routines, which calls into question their authority and credibility (Springer et al. 2015; Krajewski and Ekdale 2017). All this soon led to the understanding that citizen journalism was not going to replace the work of the media's news work but rather would occupy a secondary role (Neuberger and Nuernbergk 2010; Lacy et al. 2010; Franquet et al. 2011).

Having thus abandoned news-creation activity in the media, the main source of participation moved to news comment sections. However, experiences in this area were often not positive, either. On the one hand, existing studies show that participation in these spaces lacks deliberation (Ruiz et al. 2011; Rowe 2015; Castellano-Parra et al. 2020; Engelke 2020) and, as acknowledged by the users themselves, on many occasions, it does not meet ideals for democratic participation and is rather more linked to entertainment (Springer et al. 2015). This all leads to doubts regarding its public-sphere nature (Almgren and Olsson 2016; Manosevitch and Tenenboim 2017; Suárez-Villegas 2017).

Yet, perhaps the most dissuasive aspect and the one that has received the most attention in the recent literature is the audiences' problematic relation with the most negative ways of participation, which concerns both users (Eberwein 2019) and the media (Frischlich et al. 2019). The rules that had characterised audience participation before the Internet (identity, relevance, brevity, authority, entertainment, and civility) gave way to others, such as anonymity, confrontation, and ultra-brevity (Pastor-Pérez 2012), opening the door to uncivil discourse, hate speech (Erjavec and Kovačič 2012; Coe et al. 2014; Harlow 2015; Su et al. 2018), and all other kinds of excesses under the cloak of anonymity (Santana 2014).

Risks are not limited to a lack of civility but can also be found in media comment sections, with disinformation campaigns or attempts to influence public opinion (Braun and Eklund 2019). This has led the media to focus part of their work on controlling these comments (Wintterlin et al. 2020), given that the disrepute they generate has an impact on corporate credibility and how they drive political action (Ardèvol-Abreu et al. 2018).

As a result, many media outlets have reduced their participation sections or have entirely closed them, diverting audience participation to social media, which has become the most usual way to interact with news media content (Almgren and Olsson 2016).

However, not all aspects are negative. Citizen journalists have demonstrated initiative in tracking down their own stories, offering first-person testimonies or testimonies based on their own experiences (Reich 2008), and they help to enrich the public sphere by offering coverage of alternative issues, events, and points of view (Nah and Yamamoto 2019). In this regard, they have certainly helped to raise visibility for ordinary citizens in the media's eyes since they are not limited to the use of habitual sources (Neuberger and Nuernbergk 2010; De Keyser and Raeymaeckers 2012).

Moreover, citizen journalism has made valuable contributions to events of great social impact (Mano and Milton 2016; Konow-Lund and Olsson 2017), showing its ability to offer information in real time directly from the place where events are happening (Hänska-Ahy and Shapour 2013; Hung 2013; Barranquero-Carretero and Meda-González 2015; Moyo 2015; Mpofu 2016; Odabaş and Reynolds-Stenson 2018). For this reason, we might consider that audiences may perform "acts of journalism" (Holt and Karlsson 2014) in a complementary way to the work carried out by the news media.

### *3.3. Distrust of Journalists and Defence of Organisational Culture*

Since the beginning, journalists have viewed citizen journalism with scepticism (Singer et al. 2011) and considered it something peripheral, which in no case was comparable to their own work (Vos and Ferrucci 2018). Contributing to this mistrust was that some of these users considered that their work was opposite to that of news media, which posed a relationship in terms of confrontation (Quandt 2018).

As time passed, some of these positions have softened, and many journalists recognise the interest in user-generated content for their daily work (Suárez-Villegas 2017). Although, in general, they still do not consider this work to be journalism, they have been abandoning adversarial positions and have begun to better value the contributions made by users (Chua and Duffy 2019).

For newsroom professionals, the essence of journalism work has not changed. Journalists have a strong professional identity, which is based on belonging to self-regulating organisations whose common objective is the production of primary information (Andersson and Wiik 2013; Örnebring 2013; Vos and Ferrucci 2018). They also perceive that their main commitment is to public service (Deuze 2005), which implies, among many other professional traits, respecting plurality, verifying information, being truthful or separating facts and opinions (Suárez-Villegas 2017). And although the deliberation strategy looked attractive because we understand that the media are spaces for public discussion, professionals perceive the time devoted to interacting with audiences as a deviation from the main activity they must carry out (Lawrence et al. 2018).

Although citizen journalism and user-generated content did not question the professional identity of the journalists (Heise et al. 2014), their emergence has reinforced it (Carlsson and Nilsson 2016).

This shared position of defending organisational culture has left little room for citizen journalism or creative participation. Neither did it contribute to improving the perception of audience contributions that the tools provided to them were, in addition to other uses, a source of harassment and hectoring against journalists (Erjavec and Kovačič 2013; Gardiner 2018).

Journalists think that user-generated content needs to be reviewed by professionals to avoid bias or manipulation (Örnebring 2013). In fact, the most usual resource is to use them as representatives of the "common people," with no regard for their specialisation, knowledge, or experience (Hermans et al. 2014). Citizens are also interested in raising new topics, although there tends to be a limited number of participants, which creates fragility (Wiard and Simonson 2019).

In contrast with citizen journalists, the majority of whom have no training in this field (Kus et al. 2017), journalists defend their profession tooth and nail, which requires, in addition to professional background, a set of skills and an institutional structure (Quandt 2018).

Journalists adduce that one of the reasons limiting audience participation is that their workload prevents them from achieving truly valuable collaborations on user-generated content. Reviewing comments is also a low priority, although they openly acknowledge that they follow social media more (Pérez-Díaz et al. 2020). Some authors have identified this lack of reciprocity from journalists as precisely one of the reasons that audiences are demotivated (Lewis et al. 2014). Even though all journalists consider dialogue enriching (Suárez-Villegas 2017), the media has conditioned how these contents are produced, their context, and even the topics (Holt and Karlsson 2011).

Journalists also do not particularly appreciate the value of users' contributions in controlling the editorial content of the media. Driven by the defence of their autonomy and their social role (Pérez-Díaz et al. 2020), mistake correction buttons or user contributions to creating and revising news pieces are two of the lowest-rated options by professionals (Ramon et al. 2020), although significantly greater with young journalists.

### *3.4. Making Participation Profitable*

For news companies, adding participation has led to new tension between the two classic newsroom discourses, meaning between the journalists' professional ideal and the company's corporate efficacy and the profitability of the business model (Andersson and Wiik 2013).

Within the context of this classic duality between quality and profitability, media managers have perceived user-generated content as a way to increase the number of visits to capture their attention for longer and garner greater loyalty to the publication and, in some cases, also reduce production costs (Vujnovic et al. 2010; Manosevitch and Tenenboim 2017). As Franck (2019) summarises, media exchange information and entertainment for attention, which is, in turn, monetised via advertising. Managers also understand that opening up the door to participation can lead to reinforcement of the organisation’s cultural capital (Hellmueller and Li 2015).

As such, since the beginning, a large portion of the draw of participation for news companies was their cost-saving aspect, given that audiences willing to create content for free were added (Domingo et al. 2008). Generalising participation spaces (there were plenty of them but very limited in nature and potential) had much more to do with economic reasons than the original spirit of promoting spaces for deliberation to broaden democratic freedoms (Singer et al. 2011; Masip and Suau 2014). As such, comment sections, surveys, forums and, in general, all spaces where audiences could share their content rapidly spread (Hermida and Thurman 2008), although, like with other innovations in the sector, the professional and corporate imperative was to occupy space in the new media, without excess concern for how this objective was achieved.

The difficulty in managing user-generated content (Domingo 2011), partially in its most negative aspect, far from providing cheap labour, made clear that audience participation required a much greater amount of technical and human resources to help filter content (Lawrence et al. 2018). Participation spaces needed to be redefined, moderation needed to be increased, new tools needed to be implemented (including machine learning), and they began attempting to identify users through platform logins.

In summary, the difficulties involved in managing participation and controlling user-generated content and the resistance to sharing editorial responsibility have led the media to gradually limit their spaces for participation (Masip et al. 2019). At the same time, the unstoppable emergence of social networks naturally transferred user-generated content and audience participation to them, while the media have preferred to opt for content distribution (Westlund 2013; Hille and Bakker 2013; Peña-Fernández et al. 2016; Larrondo-Ureta et al. 2023).

This was most certainly one of the main transformations in how news media conceived audience participation. Several longitudinal studies on how media professionals perceive the contribution that their audiences might make indicate that they have gone from perceiving them as potential content co-creators to mainly considering them as re-distributors of information (Krumsvik 2018). This shift in focus on relations between players in the news process has led to a reduced number of media participation spaces and an increase in practises to draw audience attention and how to make this attention profitable (Myllylahti 2020).

The strategy of diverting participation towards social media has the advantage that it is a complementary channel for traffic and income that does not question the legitimacy of the journalist profession (Vos and Ferrucci 2018). Thus, the news media can maintain their role as a primary source of news, newsrooms maintain editorial control over the content they create, and audiences later help to raise visibility for this news, re-distributing it and interacting with it once published, which contributes to strengthening a business model that guarantees the social role of the media (Krumsvik 2018). But at the same time, this practice also delves into the antagonism between media corporations and Internet platforms to the extent that it contributes to the overproduction of content and greater relevance of the platforms’ intermediation work. This has undermined the traditional business and production model of journalism and the logic of concentration and control of production (Siapera 2013).

It is at this point that we find one of the subtle paradoxes of audience influence on news media content. Although the possibility of directly creating news in the media has flagged, the capacity to share the news and guide professional journalist routines towards more commercial, profit-oriented perspectives, has increased (Belair-Gagnon and Holton 2018).

Banking on social media, therefore, is not harmless, as it fully involves the news media in the fight for audience attention (Myllylahti 2020). Web analytics companies do not consider journalist criteria but rather indirectly promote standards and values oriented towards obtaining profits by introducing the visibility or success rates of newsrooms amongst users in said newsrooms' routine values (Belair-Gagnon and Holton 2018). In the last instance, this all flames the debate on how this attention economy jeopardises the public sphere and the very operation of democracy (Hindman 2018).

## 4. Conclusions and Discussion

The emergence of online news media, especially beginning with Web 2.0, empowered audiences as voices in parallel to the news media to create their own content through easy-to-use, low-cost tools. In some cases, this developed news utopias where people spoke of "journalism without journalists".

The promises of citizen journalism and audience participation in creating news in the media soon came up against three important limitations that stunted their development. On the one hand, the audience's lack of interest in creating news content in a sustained manner, despite the ability they have shown to make relevant and valuable contributions to events of great social impact (disasters, mass events, etc.). Likewise, the media have had difficulties in managing their audiences' oftentimes problematic participation. The redefinition of the traditional norms with which audiences had contributed until then in the media, in particular, anonymity or the volume of participation, has increased the risk of incivility and the spread of information disorders.

In addition, corporate zeal, journalists' defence of their organisational culture, and a corporate approach based on securing audience loyalty and economic profitability (closely related to plummeting advertising income and the downfall of structures that the media had on similar channels) were far from the original idea of fomenting a deliberative culture and distributing power in democratic societies.

Even though there is no doubt that citizen journalism and participation have redeemed the central role of audiences in the news process (Picone et al. 2015) and have also sparked a greater degree of participation and interactivity (Jönsson and Örnebring 2011), they do not participate in the channels created in the media (Suau et al. 2019). Neither have journalists been able to create routines and spaces for interaction with audiences in the media themselves, while companies have focused on seeking out new ways to sell their products without giving up editorial control (Usher 2014).

Due to all the aforementioned, the idea of citizen journalism offering user-generated content through the media outlets themselves now appears to be a thing of the past (Quandt 2018) and bears many of the features defining failed innovation, given that citizen journalism currently occurs on the periphery of traditional news media (Wall 2019).

However, giving up on co-creation spaces does not mean that audiences have lost their capacity to influence the content that the media create. Paradoxically, through social media, where their role as news re-distributors has granted audiences a gatekeeping role after publication (Hermida 2020), their influence has directly conditioned the content that the media create based on their success on social media. Users are advocates who redefine the hierarchy of news that was previously selected by journalists, deciding which news is worth sharing and which is not (Masip 2015).

Redirecting participation towards social media seeks to maximise media audience, impact, and revenue. Citizen journalism, or the co-creation of news content, has been diluted in favour of SEO techniques or the quasi-exclusive conception of participation, as opinions conducted on channels that prioritise audience loyalty and provide access to new audiences through channels have never before been considered incidental consumption of information (Boczkowski et al. 2018). In the classic dispute between newsroom and corporate values, audience participation appears to shift the scale towards the latter (Andersson and Wiik 2013).

As such, the result of empowering audiences who are active in the media is largely mediatised participation, leaning towards greater audience interaction with media content. The struggle to draw their interest in a highly competitive setting constitutes a new method of participation (indirect and "ex-post", although as influential) that once again places the old controversy between public interest and the interest of the public at the heart of the debate. For this reason, it is important that news companies continue to establish collaboration channels amongst all players who participate in the process (Eldridge 2018) so they can continue innovating (Westlund et al. 2021).

This study bears certain limitations. Since it exclusively considers how citizens participate in the media, it does not address how this citizen journalism has moved to social media (Ritonga and Syahputra 2019) or community or alternative media that have arisen in parallel with traditional media, oftentimes much smaller in size and focused on defending specific interests (Harcup 2011; Meadows 2013).